\documentclass[jcp,twocolumn,amsmath,amsfonts,amssymb,showpacs,final]{revtex4-1}

\usepackage{graphicx,graphicx,epsfig,amsfonts,amssymb}

\usepackage{bm}
\usepackage[all]{xy}

\newcommand{\beq}{\begin{equation}}
\newcommand{\eeq}{\end{equation}}

\newcommand{\qed}{\nobreak \ifvmode \relax \else
      \ifdim\lastskip<1.5em \hskip-\lastskip
      \hskip1.5em plus0em minus0.5em \fi \nobreak
      \vrule height0.75em width0.5em depth0.25em\fi}

\usepackage{bm}
\usepackage[all]{xy}

\begin{document}
\title{Supersymmetry and fluctuation relations for currents in closed networks}

\author {Nikolai A. {Sinitsyn}$^{a,b}$}
\author{Alexei {Akimov}$^{b,c}$}
\author{Vladimir~Y. {Chernyak}$^{a,d}$}
\address{$^a$ Theoretical Division, Los Alamos National Laboratory, B258, Los Alamos, NM 87545}
\address{$^b$New Mexico Consortium, Los Alamos, NM 87544, USA}
\address{$^c$Department of Chemistry, Rice University, Houston, Texas 77005}
\address{$^d$Department of Chemistry, Wayne State University, 5101 Cass Ave,Detroit, MI 48202}

\date{\today}

\begin{abstract}
We demonstrate supersymmetry in the counting statistics of stochastic particle currents and use it to derive exact nonperturbative relations for the  statistics of currents induced by arbitrarily fast time-dependent protocols.
\end{abstract}

\date{\today}

\maketitle

{\it Introduction}. The discovery of  fluctuation theorems and nonequilibrium work relations \cite{Evans93} has stimulated considerable interest in nonequilibrium statistical mechanics and theory of counting statistics \cite{andrieux-09njp}. Fluctuation theorems apply to evolution of thermodynamically important characteristics, such as work, dissipated heat and entropy. The utility and even a proper definition of these thermodynamic concepts in the framework of mesoscopic nonequilibrium physics are still a subject of considerable debates. Hence, it is important to obtain  exact relations that do not directly rely on the thermodynamic concepts, but rather describe unambiguous characteristics, such as statistics of particle currents in systems driven by time-dependent fields.  Progress in this direction has been relatively modest. Fluctuation theorems have led to exact relations for statistics of particle currents only in nonequilibrium {\it steady states} \cite{kurchan-ft-review,andrieux-09njp}, or for driving protocols that do not break time-reversal symmetry \cite{ft-review}.

In this letter, we present exact relations for statistics of currents in strongly driven mesoscopic stochastic systems. Being akin to known fluctuation theorems, a part of our exact result is not directly related to the condition of microscopic reversibility but rather follows from {\it supersymmetry} of the counting statistics of currents.

\begin{figure}
\centerline{\includegraphics[width=1.8in]{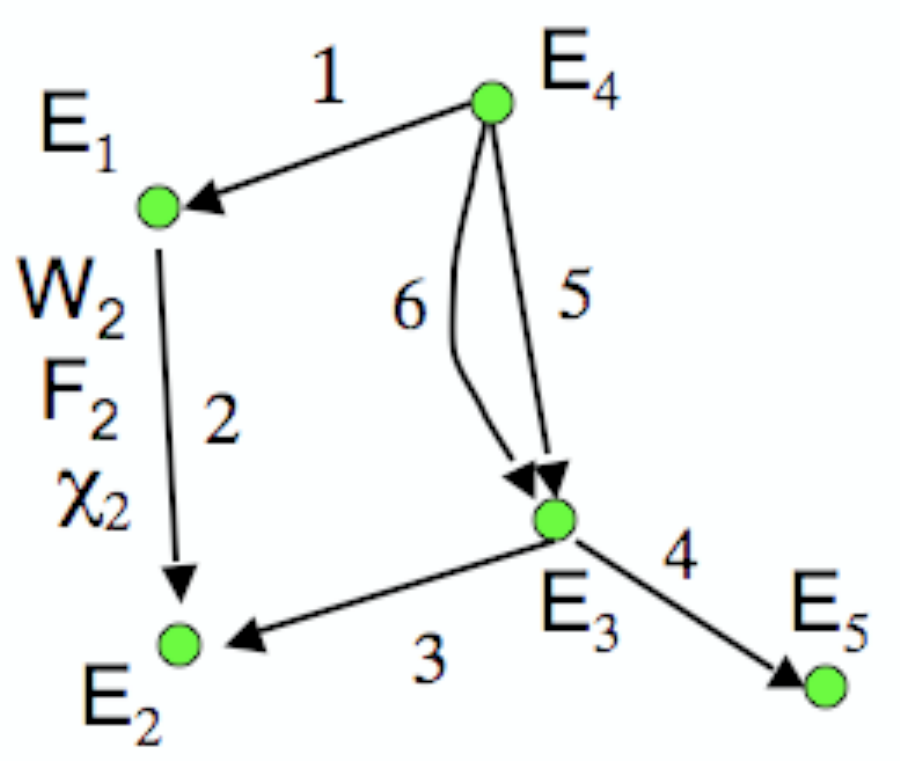}}
  \caption{ A closed graph with 5 nodes and 6 links, representing a Markov chain with five states. Links represent allowed transitions with rates
$k_{ij}^{\alpha}=ke^{\beta (E_i-W_{\alpha}+f_{ij}^{\alpha})}$.  Nodes are characterized by energies of wells $E_i$, $i=1,\ldots , 5$.
 Links are characterized by sizes of barriers $W_{\alpha}$, by parameters $F_{\alpha} \equiv F^{\alpha}_{ij}=ln(k^{\alpha}_{ij}/k^{\alpha}_{ji})= f_{ij}^{\alpha}-f_{ji}^{\alpha}+E_i-E_j$, and
by counting parameters $\chi_{\alpha} \equiv \chi_{ij}^{\alpha}$ that are used to count number of transitions through the link $\alpha$.
Here $\alpha=1,\ldots, 6$.}
\label{network}
\end{figure}

{\it Model.} In this section, we introduce a notation that will be used to formulate our results. Consider a graph $X=(X_{0},X_{1})$, with $X_{0}$ and $X_{1}$ being the node and link sets, respectively, and a Markov process  describing a particle motion on $X$ with a set of transition rates, $k_{ij}^{\alpha}$, of jumping from the node $i$ to the node $j$ through the link $\alpha$. Each link connects two different nodes. Fig.~\ref{network} shows an example of such a graph with 6 links and 5 nodes. To define positive direction of currents, we prescribe arbitrary orientations on all links. We denote by $C_{0}(X)$ and $C_{1}(X)$ the vector spaces of distributions ${\bm\rho}=(\rho_{j}|j\in X_{0})$ and currents on links ${\bm J}=(J_{\alpha}|\alpha\in X_{1})$, respectively. Nodes and links are labeled by Latin and Greek indices, respectively. We also introduce a notation, $\partial\alpha=(i,j)=(\partial_{0}\alpha,\partial_{1}\alpha)$, which means that $\partial\alpha$ is the oriented border of the link $\alpha$, and it can be represented by an ordered pair of nodes $(i,j)$ where $j$ is a node to which the link $\alpha$ points and $i$ is the node from which this link originates. With a minimal abuse we will use the same notation, $\partial\alpha=\{i,j\}$, for non-oriented boundary. It is convenient to view a current as a set of components $(J_{ij}^{\alpha})$, with $\{ i,j \}=\partial\alpha$, such that $J_{ij}^{\alpha}=-J_{ji}^{\alpha}$. Conservation of particles requires that $\sum_j \rho_j =1$.
The evolution of the probability vector is given by the {\it master equation},
\begin{equation}
d_{t}{\bm\rho}=\hat{H}{\bm\rho},
\label{master}
\end{equation}
where we will call $\hat{H}$ the {\it master operator}. Let $\vert j \rangle$ and $\langle i \vert$ be the bra- and ket-vectors over the space $C_0(X)$ with the only nonzero unit components at $j$-th and $i$-th positions respectively. In this basis set,
$ H_{ij}=\sum_{\alpha}^{\partial \alpha=\{i,j\}}k_{ji}^{\alpha}$ for $i\ne j$, and $H_{ii} = -\sum_{\alpha,j}^{\partial\alpha=\{i,j\}}k_{ij}^{\alpha}$.

Detailed balance (DB) is not assumed, the deviation from DB is quantified by the {\it  entropy
function} (EF) that associates $F({\bm l})\equiv\ln\prod_{\alpha \in {\bm l}}(k_{ij}^{\alpha}/k_{ji}^{\alpha})$ with a closed path ${\bm l}$, so that, for time-independent parameters, $e^{F({\bm l})}$ represents the ratio of the probability of a stochastic trajectory ${\bm l}$ to its time-reversed counterpart. Here we introduced a notation, $\prod_{\alpha \in {\bm l}} (k_{ij}^{\alpha}/k_{ji}^{\alpha})$, which means the product of $k_{ij}^{\alpha}/k_{ji}^{\alpha}$ over all links of the loop, where we assume an arbitrary direction of motion along the loop as positive and  take $(i,j)=\partial \alpha$ if the link $\alpha$ points along the positive loop direction and $(j,i) =\partial \alpha$ if the link points against direction of the loop. Later, we will use analogous notation for the sum over loop links, $\sum_{\alpha \in {\bm l}}$. Note that DB means that $F({\bm l})=0$ for any closed path ${\bm l}$. We will consider only driving protocols that conserve the entropy functions $F({\bm l})$ at any cycle ${\bm l}$ of a graph.


We focus on the case of periodic driving, when the kinetic rates depend on time in a periodic manner, according to a driving protocol with the time period $\tau$. Such a steadily driven system eventually enters a regime with periodically changing population probability vector, i.e.  ${\bm\rho}(t+\tau)={\bm\rho}(t)$.
We can introduce the currents per period of the driving protocols ${\bm J^c}=(\int_0^{n\tau}dt{\bm J}(t))/n$. Since, in the limit $n\rightarrow \infty$, particles cannot accumulate in any node, the sum of currents entering any node is equal to the sum of the currents leaving this node. We will call  currents with this property the {\it conserved currents}.

The entropy function is naturally extended  to a linear entropy functional on currents, given by
\begin{equation}
{\cal E}({\bm J^c})=\sum_{\alpha\in X_{1}}F_{\alpha}J^c_{\alpha},
\label{entropy}
\end{equation}
where we introduced the vector ${\bm F}$, whose components are indexed by the links, given by
$F_{\alpha} \equiv F_{ij}^{\alpha}=-F_{ji}^{\alpha}\equiv\ln(k_{ij}^{\alpha}/k_{ji}^{\alpha})$. It is important to note that even though the parameters $k_{ij}^{\alpha}$, and hence $F_{ij}^{\alpha}$, are time-dependent, the entropy functional ${\cal E}$ defined on conserved currents ${\bm J^c}$ is time-independent, provided all $F({\bm l})$ are time-independent. One can see this, e.g., by noticing that there is a basis set in the space of possible conserved currents, which consists of constant unit-valued currents circulating in each independent cycle of a graph and having zero values on all other links. The contribution of each such independent conserved current to the entropy functional is just the entropy function for a corresponding loop of a graph, and any conserved current is just a linear combination of  such circulating basis currents.

{\it Fluctuation relation for currents} (FRC). Consider a Markovian kinetics of a particle on a graph, with constant entropy function, $F({\bm l})$, and periodically time-dependent kinetic rates, $k_{ij}^{\alpha}$. The probability distribution of conserved currents, generated per period of driving, has the large-deviation  (LD) form $P({\bm J^c},n)\sim e^{n{\cal S}({\bm J^c})}$, with ${\cal S}({\bm J^c})$ being referred to as the LD function. Consider the  following symmetry property of the LD function ${\cal S}({\bm J^c})$ of the  conserved currents
\begin{eqnarray}
\label{current-fluct-theor} {\cal S}^{f}({\bm J^c})-{\cal S}^{b}(-{\bm J^c})={\cal E}({\bm J^c}),
\end{eqnarray}
where $f$ and $b$ stand for the original (forward) and time-reversed (backward) driving protocols respectively. Kinetic rates in backward and forward protocols are related by $k_{ij}^{\alpha, b}(t)=k_{ij}^{\alpha, f}(\tau-t)$. Eq.~(\ref{current-fluct-theor}) is known to hold if the kinetic rates are independent of
time
\cite{kurchan-ft-review,ft-review} but it does not hold for general time-dependent rates \cite{ft-review}.  Nevertheless we can show, and this is the content of our FRC, that Eq.~(\ref{current-fluct-theor}) does hold for two types of periodic driving protocols, for which, in addition to conservation of entropy function, either of the two conditions is satisfied:

(i)  the ratios $k_{ij}^{\alpha}/k_{ji}^{\alpha}$ of the forward/backward rates are kept time-independent at all links $\alpha$. This can be described by introducing time dependent parameters, $W_{\alpha}$, on the links, such that   $k_{ij}^{\alpha}(t)=\bar{k}_{ij}^{\alpha}e^{W_{\alpha}(t)}$ .

(ii) the branching ratios $k_{ij}^{\alpha}/k_{ij'}^{\alpha'}$ are kept time-independent at all nodes. This can be described by introducing time-dependent parameters, $E_{i}$, on nodes, such that  $k_{ij}^{\alpha}(t)=\widetilde{k}_{ij}^{\alpha}e^{-E_{i}(t)}$.

It is possible to write kinetic rates in the form $k_{ij}^{\alpha}=e^{W_{\alpha}-E_{i}+f_{ij}^{\alpha}}$, with a constant vector  ${\bm f}=(f^{\alpha}_{ij}|\alpha \in X_{1})$ on the links, and, generally,
$f^{\alpha}_{ij} \ne f^{\alpha}_{ji}$ . For the case of a network that describes transitions of a physical system among deep free energy minima in its phase space, parameters $W_{\alpha}$, $E_{i}$, and  $f_{ij}^{\alpha}$ have a clear physical interpretation \cite{jarzynski-08prl}. They correspond, respectively, to
the potential barrier separating metastable states along the path $\alpha$, to the size of the energy of a well in the node $i$, and to the effect of an external force acting on the system along the path $\alpha$. Adopting this terminology, we can formulate the FRC as the statement of the validity of Eq.~(\ref{current-fluct-theor}) for time-dependent protocols in which either (i) only potential barriers are driven or (ii) only node energies  are driven.

It is important to note that parameterization of kinetic rates by the set $({\bm E},{\bm W}, {\bm f})$ is not unique since it is possible to redefine parameters $f_{ij}^{\alpha} \rightarrow f_{ij}^{\alpha}-w_{\alpha} - \varepsilon_{i}$, $W_{\alpha} \rightarrow W_{\alpha}+w_{\alpha}$, and $E_i \rightarrow E_i-\varepsilon_i$, and obtain the same set of kinetic rates $k^{\alpha}_{ij}$. This change of parameters preserves the entropy function. In fact, a set of kinetic rates is fully determined by the entropy function $F({\bm l})$ and the sets ${\bm E}=(E_{j}|j\in X_{0})$ and ${\bm W}=(W_{\alpha}|\alpha\in X_{1})$. The entropy functions for any loop ${\bm l}$ can then be expressed in terms of ${\bm f}$ alone, i.e. $F({\bm l})=\sum_{\alpha \in {\bm l}} (f_{ij}^{\alpha}-f_{ji}^{\alpha})$; this means that the entropy function and the entropy functional (\ref{entropy}) are invariant with respect to the rate transformations $k_{ij}^{\alpha}\mapsto k_{ij}^{\alpha}e^{w_{\alpha}-\varepsilon_{i}}$. We will use this property in our derivation of FRC, namely, if we can prove FRC for some choice of the vector ${\bm f}$ which corresponds to a given form of the entropy functional, then FRC is valid for any other choice of a constant ${\bm f}$ that corresponds to the same entropy functional. During the derivations of (i) and (ii), different choices of ${\bm f}$ will be used to apply the symmetries of the problem.

{\it Operator derivation of case (i) and Lagrangian interpretation.} We will derive FRC by considering the
symmetries of the Legendre transform, $\omega_{\bm \chi}$, of ${\cal S}({\bm J^c})$, where $\bm \chi$ is the variable conjugated to ${\bm J^c}$. In the literature, $\omega_{\bm \chi}$ is often called the  cumulant generating function of currents \cite{sukhorukov-07Nat}. An operator approach to deriving the cumulant generating function of conserved currents is based on introducing the {\it twisted master operator}, $\hat{H}_{\bm\chi}$, parameterized by the multi-variable argument ${\bm\chi}=(\chi_{\alpha}|\alpha\in X_{1})$ (or alternatively by a set of antisymmetric components $\chi_{ij}^{\alpha}=-\chi_{ji}^{\alpha}$, called counting parameters) of the generating function. Here the word ``twisted'' reflects the way the operator $\hat{H}_{\bm\chi}$ is constructed, i.e. by multiplying (twisting) off-diagonal elements of $\hat{H}$ by corresponding $e^{\chi_{ij}^{\alpha}}$ factors, namely, the operator $\hat{H}_{\bm\chi}$ is obtained by replacing the off-diagonal components of $\hat{H}$ with $\langle i|\hat{H}_{{\bm\chi}}|j\rangle=\sum_{\alpha}^{\partial\alpha =\{i,j\}}e^{\chi_{ji}^{\alpha}}k_{ji}^{\alpha}$, and with the diagonal components remaining the same \cite{sukhorukov-07Nat}.

After many periods of driving, $\omega_{{\bm\chi}}$ is determined by the largest eigenvalue, $e^{\omega_{{\bm\chi}}}$,  of the evolution operator
\begin{eqnarray}
\label{evolve-period} \hat{U}_{\bm\chi}\equiv \hat{T}\exp\left(\int_{0}^{\tau}\hat{H}_{{\bm\chi}}(t)\right), \quad \hat{U}_{\bm\chi}|\psi_{{\bm\chi}} \rangle=e^{\omega_{{\bm\chi}}}|\psi_{{\bm\chi}}\rangle.
\end{eqnarray}
Eq.~(\ref{current-fluct-theor}) is then equivalent to the following symmetry property of $\omega_{{\bm\chi}}$:
\begin{eqnarray}
\label{current-fluct-theor-generating} \omega_{{\bm\chi}}^{f}=\omega_{-{\bm\chi}-{\bm f}}^{b}.
\end{eqnarray}

For case (i), Eq.~(\ref{current-fluct-theor-generating}) can be derived in a concise way. By direct verification, we have the symmetry
\begin{equation}
\hat{H}_{{\bm\chi}}^{T}=\hat{H}_{-{\bm\chi}-{\bm F}},
\label{symmetry}
\end{equation}
where $\hat{H}_{{\bm\chi}}^{T}$ is the transpose of $\hat{H}_{{\bm\chi}}$ and, by the condition imposed on the driving protocol, (i), the vector ${\bm F}$ is time-independent. The transposition changes the ordering of operator products, while maintaining the eigenvalues unchanged. This implies $(\hat{U}_{{\bm\chi}}^{f})^{T}=\hat{U}_{-{\bm\chi}-{\bm F}}^{b}$. Eq.~(\ref{current-fluct-theor-generating}) follows from the fact that for a constant set ${\bm E}$ we can always redefine the set ${\bm f}$ so that ${\bm f}={\bm F}$.

Our elementary derivation of (i) is complemented by a simple interpretation in terms of probabilities of particle trajectories, $P({\bm x})$, where a closed particle trajectory ${\bm x}=({\bm l},{\bm t})$ is represented by a closed path ${\bm l}$ on $X$, and temporal data ${\bm t}$ (the times when the jumps occurred). We then have the symmetry property $P^{f}({\bm x})/P^{b}({\bm x}^{r})=e^{F({\bm l})}$ for the probabilities of the original trajectory and time-reversed trajectory ${\bm x}^{r}$ for the time-reversed protocol. For case (ii), the above arguments cannot be applied, since ${\bm F}$ depends on energies, which in case (ii) are time-dependent, so that the symmetry (\ref{symmetry}) cannot be considered equivalent to
(\ref{current-fluct-theor-generating}). Neither does case (ii)  have a simple interpretation in terms of stochastic trajectories. We will derive Eq.~(\ref{current-fluct-theor}) for case (ii) using an additional supersymmetry property of the evolution with twisted master operator.




\begin{figure}
\centerline{\includegraphics[width=2.8in]{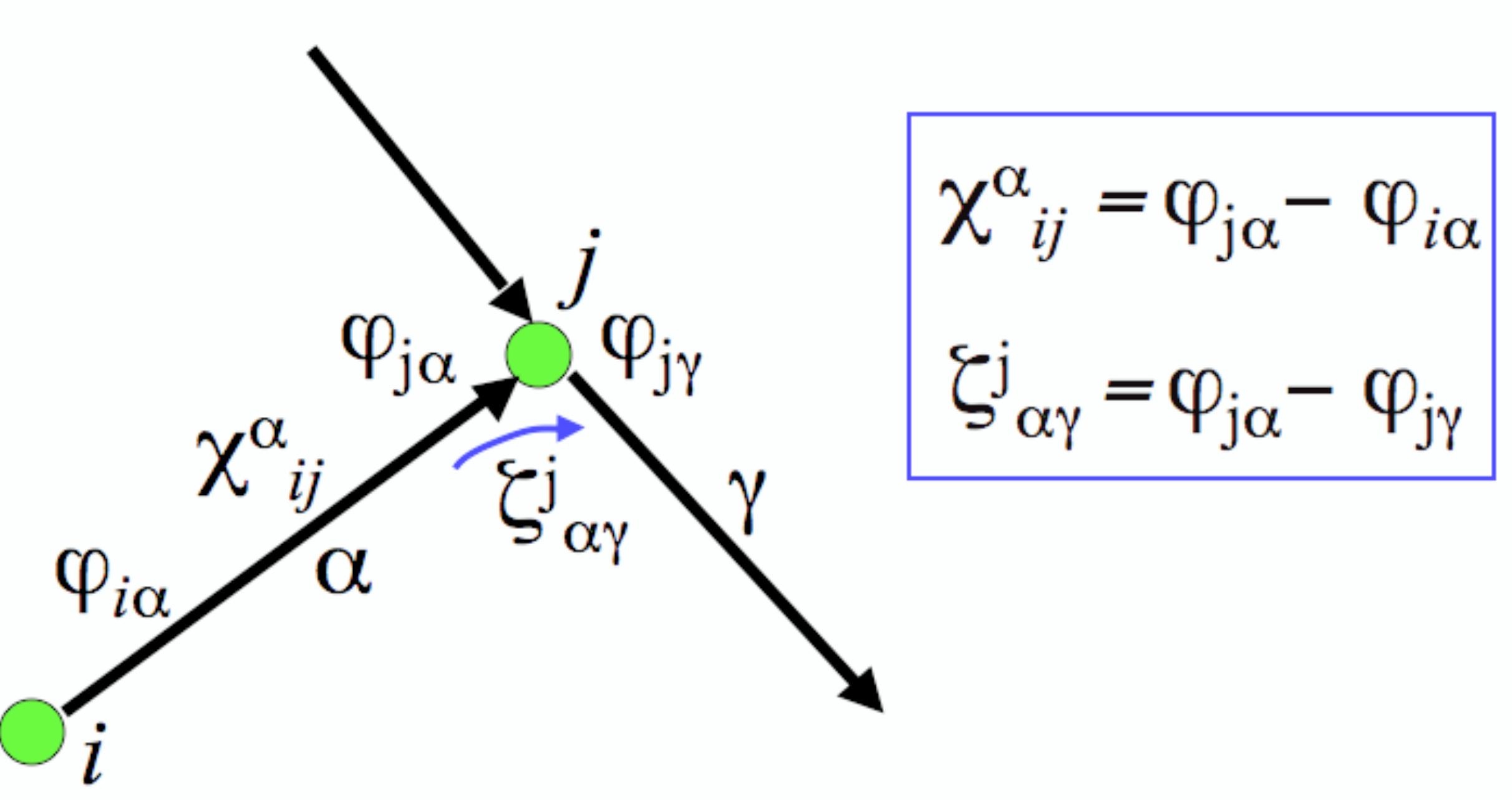}}
  \caption{ Equivalence relations among counting parameters.}
\label{parameters}
\end{figure}

{\it Supersymmetry for master equation}. A hidden supersymmetry of Langevin dynamics and motion on cyclic graphs has been discussed  in the literature (see e.g. in \cite{duality1}). Here we show that a hidden supersymmetry can be found in Markovian evolution on an arbitrary graph. More importantly, this supersymmetry can be extended to make it a property of the counting statistics of currents. The master operator has a representation $\hat{H}=-{\cal Q}\hat{J}$, with $\hat{J}$  being the current operator, and $\hat{J} \vert i \rangle = \left(\sum_{\alpha,j}^{\partial \alpha = (j,i)} k_{ij}^{\alpha} -\sum_{\alpha,j}^{\partial \alpha = (i,j)} k_{ij}^{\alpha} \right) \vert \alpha \rangle $. The operator ${\cal Q}$ acts from $C_{1}$ to $C_{0}$ by ${\cal Q}|\alpha\rangle\equiv|j\rangle-|i\rangle$ for $\partial\alpha=(i,j)$. Here $\vert \alpha \rangle$ is the vector in $C_{1}$ with a unit entry corresponding to link $\alpha$ and zero otherwise.  The operator ${\cal Q}$ plays the role of a discrete counterpart of the ${\rm div}$-operator \cite{jarzynski-08prl}. Therefore, for conserved currents, we have ${\cal Q}{\bm J^c}=0$.

We further introduce an operator $\widetilde{H}=-\hat{J}{\cal Q}$ that acts in the space $C_{1}$ of currents, and notice that obviously $\hat{H}{\cal Q}={\cal Q}\widetilde{H}$. Viewing $\bar{H}\equiv \hat{H}\oplus\widetilde{H}$ and ${\cal Q}$ as operators acting in $\bar{C}\equiv C_{0}\oplus C_{1}$, we have $[{\cal Q},\bar{H}]=0$. Considering $C_{0}$ and $C_{1}$ as even and odd components,
of the (super) space, ${\cal Q}$ becomes an odd operator that commutes with the master (super) operator $\bar{H}$, and $\bar{H}$ can be written as the anticommutator of $-{\cal Q}$ and $\hat{J}$, which closes the algebra; therefore, the term supersymmetry is totally appropriate here. Supersymmetry connects evolution of the probability distribution  and current distribution components
of the superspace, $\bar{C}$, thus allowing the standard master equation to be reformulated in terms of its counterpart that describes evolution in the space of currents. Our key observation for proving (ii) is that supersymmetry for master operator also holds for the twisted operator, $\hat{H}_{{\bm\chi}}$. By reformulating the problem of finding the generating function in terms of the superpartner of $\hat{H}_{{\bm\chi}}$, we will show that a derivation of case (ii) becomes as simple as for case (i).




{\it Supersymmetry for twisted operators.} To identify supersymmetry on the level of twisted operators we need to come up with a procedure for twisting the current master operator $\widetilde{H}$. Since the off-diagonal elements of $\widetilde{H}$ are between the links that share a common node, it is reasonable to represent the twisting data by ${\bm\zeta}=(\zeta_{\alpha\gamma}^{j}|j\in\partial\alpha\cap\partial\gamma)$ with antisymmetric components $\zeta_{\alpha\gamma}^{j}=-\zeta_{\gamma\alpha}^{j}$. This allows a family $\widetilde{H}_{{\bm\zeta}}$ of twisted current master operators to be introduced by
\begin{eqnarray}
\label{twist-master-odd} \widetilde{H}_{{\bm\zeta}}|\alpha\rangle &\equiv& -(k_{ji}^{\alpha}+k_{ij}^{\alpha})|\alpha\rangle   \nonumber \\  &+&
( \sum_{m,\gamma \ne \alpha}^{\partial\gamma=(j,m)} e^{\zeta_{\alpha\gamma}^{j}}k_{jm}^{\gamma}- \sum_{m,\gamma \ne \alpha}^{\partial\gamma=(m,j)} e^{\zeta_{\alpha\gamma}^{j}}k_{jm}^{\gamma}
 \nonumber \\ &+&
\sum_{m,\gamma \ne \alpha}^{\partial\gamma=(m,i)}e^{\zeta_{\alpha\gamma}^{i}}k_{im}^{\gamma} -\sum_{m,\gamma \ne \alpha}^{\partial\gamma=(i,m)}e^{\zeta_{\alpha\gamma}^{i}}k_{im}^{\gamma} )|\gamma\rangle,
\end{eqnarray}
for $\alpha=(i,j)$, as well as a family $\widetilde{U}_{{\bm\zeta}}$ of twisted evolution operators, using a definition, similar to Eq.~(\ref{evolve-period}).

Parameters, ${\bm \chi}$ and $ {\bm \zeta}$,  have a purpose of counting how many times a particle passes, respectively, through links and nodes of a graph along specified directions. The number of independent
conserved currents, however, is smaller than the sizes of these vectors.  The information about conserved currents is contained in dependence of the generating function only on expressions, $\Gamma_{{\bm\chi}}({\bm l})\equiv \sum_{\alpha \in {\bm l}}\chi_{ij}^{\alpha}$ or $\Gamma_{{\bm\zeta}}( {\bm l})  \equiv \sum_{j \in {\bm l}}\zeta_{\alpha\gamma}^{j}$, where the last sum runs over nodes that belong to the loop ${\bm l}$, and indexes $\alpha$ and $\gamma$ in $\zeta_{\alpha\gamma}^{j}$ correspond to the links that,  respectively, precede  and follow the node $j$ along the positive loop direction.
We will call two  sets, ${\bm\chi}$ and ${\bm\zeta}$, equivalent, if $\Gamma_{{\bm\chi}}({\bm l})=\Gamma_{{\bm\zeta}}({\bm l})$ for any closed path, ${\bm l}$. A simple, yet important observation is that ${\bm\chi}$ is equivalent to ${\bm\zeta}$ if and only if there is a set ${\bm\varphi}=(\varphi_{j\alpha}|j\in\partial\alpha)$ such that $\chi_{ij}^{\alpha}=\varphi_{j\alpha}-\varphi_{i\alpha}$ and $\zeta_{\alpha\gamma}^{j}=\varphi_{j\alpha}-\varphi_{j\gamma}$. We additionally illustrate the meaning and relations among the counting parameters in Fig.~\ref{parameters}. It is straightforward to verify that, provided ${\bm\varphi}$ establishes equivalence between ${\bm\chi}$ and ${\bm\zeta}$, the following supersymmetry relation takes place:
\begin{eqnarray}
\label{supersymmetry-twisted} \hat{H}_{{\bm\chi}}{\cal Q}_{{\bm\varphi}}={\cal Q}_{{\bm\varphi}}\widetilde{H}_{{\bm\zeta}}, \;\;\; \hat{U}_{{\bm\chi}}{\cal Q}_{{\bm\varphi}}={\cal Q}_{{\bm\varphi}}\tilde{U}_{{\bm\zeta}},
\end{eqnarray}
where $\cal{Q}_{{\bm\varphi}}$ is the twisted supersymmetry operator, defined by ${\cal Q}_{{\bm\varphi}}|\alpha\rangle=e^{\varphi_{j\alpha}}|j\rangle-e^{\varphi_{i\alpha}}|i\rangle$ for a link $\alpha=(i,j)$. Although Eq.~(\ref{supersymmetry-twisted}) can be verified directly by inspecting the matrix elements, it is instructive to note that it follows immediately from the representations $\hat{H}_{\bm\chi}=-{\cal Q}_{{\bm\varphi}}\hat{J}_{{\bm\varphi}}$, and $\widetilde{H}_{\bm\zeta}=-\hat{J}_{{\bm\varphi}}{\cal Q}_{{\bm\varphi}}$, where $\hat{J}_{{\bm\varphi}}$ is the twisted current operator, given by $\hat{J}_{{\bm\varphi}}|i\rangle=
\left(\sum_{\alpha,j}^{\partial \alpha = (j,i)} k_{ij}^{\alpha}e^{-\varphi_{i\alpha}} -\sum_{\alpha,j}^{\partial \alpha = (i,j)} k_{ij}^{\alpha} e^{-\varphi_{i\alpha}} \right) |\alpha\rangle$.
We note also that ${\cal Q}_{{\bm\varphi}}$ is time-independent so that the evolution operators with $\hat{U}_{{\bm \chi}}$ and $\widetilde{U}_{{\bm\zeta}}$ have the same sets of nonunit  eigenvalues even if the parameters ${\bm E}$ are periodically driven since if $|\widetilde{\psi}\rangle $ is the eigenstate of the $\widetilde{U}_{{\bm\zeta}}$, then ${\cal Q}_{{\bm\varphi}} |\widetilde{\psi}\rangle$ is the eigenstate of $\hat{U}_{{\bm \chi}}$.

{\it Derivation of case (ii)}.
To reboot the operator derivation of Eq.~(\ref{current-fluct-theor-generating}) for case (ii), we introduce the superpartner, $\widetilde{{\bm F}}=(\widetilde{F}_{\alpha\gamma}^{j}|j\in\partial\alpha\cap\partial\gamma)$, of ${\bm F}$ with the antisymmetric components $\widetilde{F}_{\alpha\gamma}^{j}=-\widetilde{F}_{\gamma\alpha}^{j}\equiv -\ln(k_{ji}^{\alpha}/k_{jm}^{\gamma})$. Obviously, the entropy functions satisfy $F({\bm l})=\widetilde{F}({\bm l})\equiv\sum_{j \in {\bm l}}\widetilde{F}_{\alpha\gamma}^{j}$ for any closed path ${\bm l}$, referred to as consistency. This means that $ \widetilde{{\bm F}}$ represents a vector which can be chosen to be equivalent to ${\bm f}$. One verifies directly that the operator (\ref{twist-master-odd}) satisfies the relation, $\widetilde{H}_{{\bm\zeta}}^{T}=\widetilde{H}_{-{\bm\zeta}-\widetilde{{\bm F}}}$,
and note that in case (ii) the data $\widetilde{{\bm F}}$ is time-independent, which results in $(\widetilde{U}_{{\bm\zeta}}^f)^{T}=\widetilde{U}_{-{\bm\zeta}-\widetilde{{\bm F}}}^{b}$ and further in the symmetry relation $\widetilde{\omega}_{{\bm\zeta}}^{f}=\widetilde{\omega}_{-{\bm\zeta}-\widetilde{{\bm F}}}^{b}$ for the eigenvalues of the evolution operators in the current space. We further naturally choose ${\bm\zeta}$ equivalent to ${\bm\chi}$; this results in $-{\bm\zeta}-\widetilde{{\bm F}}$ being equivalent to $-{\bm\chi}-{\bm f}$, due to consistency between ${\bm f}$ and $\widetilde{{\bm F}}$. This allows us to apply supersymmetry [Eq.~(\ref{supersymmetry-twisted})], which results in $\omega_{{\bm\chi}}^{f}=\widetilde{\omega}_{{\bm \zeta}}^{f}$ and $\omega_{-{\bm \chi}-{\bm f}}^{b}=\widetilde{\omega}_{-{\bm \zeta}-\widetilde{{\bm F}}}^{b}$, and further in Eq.~(\ref{current-fluct-theor-generating}), which completes our derivation of FRC.

{\it Discussion}. The FRC is applicable to the case when either energies of discrete states or the heights of barriers that separate the states are varied. This regime can be realized for catenane molecular motors or electric circuits in incoherent regime, when the gate voltages of the quantum dots are varied in time \cite{jarzynski-08prl}. Statistics of currents can be probed in single molecule measurements  \cite{single-mol-book} or in experiments with nanoscale electric circuits \cite{sukhorukov-07Nat}. 
The FRC demonstrates that there are fluctuation theorems that do not directly follow from the relations between the probabilities of forward and time-reversed trajectories. We expect that the FRC is only one of many possible applications of the supersymmetry of the counting statistics of currents.

\section*{Acknowledgment}
We are grateful to Misha Chertkov, Jordan Horowitz and Allan Adler for useful discussions. {\it This material is based upon work supported by the National Science Foundation under CHE-0808910 at WSU, and under ECCS-0925618 at NMC. The work at LANL was carried out under the auspices of the National Nuclear Security Administration of the U.S. Department of Energy at LANL under Contract No. DE-AC52-06NA25396.}


\end{document}